# Existence of Discrete Traveling Waves in Fully Coupled Network of Mackey-Glass Relay Generators

V. Alekseev (http://orcid.org/0000-0001-9723-7802) [a, *],
M. Preobrazhenskaia (http://orcid.org/0009-0000-5974-6614) [a],
V. Zelenova (http://orcid.org/0009-0000-5974-6614) [a]

[a]Centre of Integrable Systems, P.G. Demidov Yaroslavl State University, Yaroslavl, Russia

*e-mail: vladislav.alexeev.yar@gmail.com

**Abstract** — A fully coupled network of Mackey-Glass generators is considered. Each generator is described by a limit equation for the Mackey-Glass equation. The right parts are represented by a relay function obtained when the exponent in the denominator of the nonlinearity tends to infinity. Discrete traveling waves are sought in the system. These modes are such that all components are represented by the same periodic function with successive (multiple of the same value) shifts. Moreover, this periodic function has the smallest number of switchings, that is, points at which the right-hand sides change the analytical form. It is shown that discrete traveling waves coexist in the system. Moreover, their number is equal to the factorial of the number of generators.

## 1. INTRODUCTION

The Mackey-Glass equation is a very well-celebrated equation that has been studied by many authors from a wide range of areas of mathematical sciences, including Nonlinear Analysis and Mathematical Biology and has a plethora of applications in Nonlinear Dynamics and other applied fields of mathematics. It first appeared as a variation model in the relative quantity of mature cells in the blood in [1]. Afterwards, it has been studied by many authors [2-7]. In particular, this equation and its modifications were used in the description of electric generators [8-12], and in simulation of chaotic signals [13-16].

The present work continues a previous research of Mackey-Glass generators [17-19]. The electric generator described by the Mackey–Glass equation [1, 20]

$$\frac{dV}{dt} = -bV + \frac{acV(t-\tau)}{1+\bigl(cV(t-\tau)\bigr)^\gamma}. \qquad (1)$$

is called a *Mackey-Glass generator* [8, 21, 22]. Here, $V(t)$ is a voltage function, $a > 0$ is the nonlinearity saturation level, $b > 0$ is the $RC$ constant, $\tau > 0$ is the time delay, the parameter $\gamma > 0$ determines the shape of the nonlinear function, and $c > 0$ is the feedback strength.

The study of systems corresponding to circuits of Mackey-Glass generators [8, 18, 21, 22] is of great interest. In [21] the dual synchronization of chaos in two pairs of one-way-coupled Mackey-Glass electronic circuits with time-delayed feedback is demonstrated experimentally and numerically. The system under consideration has two drive generators, two response generators and feedback loop. The authors of [22] investigated the stability and occurrence of Andronov-Hopf bifurcation by analyzing the distribution of the roots of the associated characteristic equation. Using the normal form theory and center manifold argument, they derived an explicit algorithm for determining the direction of the Andronov-Hopf bifurcation and the stability of the bifurcating periodic solutions.

A similar system of Mackey-Glass generators, but for two equations and with a slightly different coupling, was considered in [8]. The model is a Mackey-Glass electronic circuit with one drive generator, two response generator and with two time-delayed feedback loops. The authors experimentally and numerically investigated chaos generation in the model. It was shown that the ratio of the two time delays is crucial for enhancing or suppressing the chaotic dynamics. It was also shown that high-quality chaos synchronization can be achieved with high coupling strengths and parameter-matching conditions between the two electronic circuits.

Now, consider an arbitrary (fixed) number of generators. There are two natural structures for coupling generators: a ring (Fig. 1) and a fully coupled network, where each is coupled to each (Fig. 2). These are solutions such that all components are represented by the same function with multiple shifts (a formal description of such solutions will be given in the next section). The technique for searching for traveling waves and proving stability in ring network is described in [23], and for a fully coupled network in [24].

In [18], a model of a ring network of $m$ Mackey-Glass generators was introduced, each of which is both drive and response, while the coupling with the previous generator has a delay:

$$\frac{dV_j}{dt} = -bV_j + \frac{ac\left(V_j(t-\tau_1) + V_{j-1}(t-\tau_2)\right)}{1 + \left(c\left(V_j(t-\tau_1) + V_{j-1}(t-\tau_2)\right)\right)^\gamma}, \quad V_0 \equiv V_m, \quad j = 1, \ldots, m. \quad (2)$$

Here, $V_j(t)$ is the voltage function in the Mackey-Glass circuit, the parameters $a, b, c, \gamma$ have the same meaning as in system (1), $\tau_1, \tau_2 > 0$ are the time delays.

In [18], for (2), the existence of modes in the form of a discrete wave is shown. In [19], for the case $m = 2$, an antiphase mode was found.

In this article we consider a fully coupled network of Mackey-Glass generators:

$$\frac{dV_j}{dt} = -bV_j + \frac{ac\left(V_j(t-\tau) + \sum_{k=0, k\neq j}^{m} V_k(t)\right)}{1 + \left(c\left(V_j(t-\tau) + \sum_{k=0, k\neq j}^{m} V_k(t)\right)\right)^\gamma}, \quad j = 0, 1, \ldots, m. \quad (3)$$

As in [17-19, 25], we will replace the system of differential-difference equations with a limit object. The limiting system has a relay right-hand side (see, for example, [30, 31]) and, at the same time, has rich dynamics and can be considered as an independent model of a generator network. Due to the fully coupled network, a large number (factorial of the number of generators in the network) of coexisting modes arises, the construction of which is the subject of this paper.

In Section 2, we introduce the object of our study: a system of Mackey-Glass equations and its limit analogue; we describe the concept of discrete travelling waves. In Section 3, we analyze the auxiliary equation with multiple delays; we define the parameters domain, for which there exists a periodical solution. In Section 4, we prove the existence of a periodical solution to the limit system in the form of a discrete travelling wave. We conclude with some numerical analysis results.

## 2. PROBLEM STATEMENT

After substitutions $V_j = c^{-1} u_j\left(\frac{t}{\tau}\right)$, $\beta = b\tau$, $\alpha = ac\tau$, and time normalization $t \mapsto \frac{t}{\tau}$, the system (3) takes a form:

$$\dot{u}_j = -\beta u_j + \frac{\alpha \left(u_j(t-1) + \sum_{k=0,k\neq j}^{m} u_k(t)\right)}{1 + (u_j(t-1) + \sum_{k=0,k\neq j}^{m} u_k(t))^\gamma}, \quad j = 0,1,\ldots,m, \quad \alpha, \beta > 0 \tag{4}$$

As in the works [18, 19, 25], we will consider $\gamma \gg 1$ as a large parameter. Denote as $F(w)$ a limit function for $\frac{1}{1+w^\gamma}$ as $\gamma \to +\infty$:

$$F(w) = \lim_{\gamma \to +\infty} \frac{1}{1+w^\gamma} = \begin{cases} 1, & 0 < w < 1 \\ \frac{1}{2}, & w = 1, \\ 0, & w > 1. \end{cases} \tag{5}$$

In the present work we obtain analytic solutions in a form of a discrete travelling wave for the limiting system of equations. Also we show a corresponding numerical solution, which agrees with the analytic results for the limiting object.

We consider a limiting (as $\gamma \to +\infty$) system of equations for a system (4)

$$\dot{u}_j = -\beta u_j + \alpha \left(u_j(t-1) + \sum_{k=0,k\neq j}^{m} u_k(t)\right) F\left(u_j(t-1) + \sum_{k=0,k\neq j}^{m} u_k(t)\right), \quad j = 0,1 \ldots, m. \tag{6}$$

We seek periodical solutions of (6) in the form of a discrete travelling wave.

### 2.1. Discrete travelling waves

Let $\pi = (k_0, k_1, \ldots, k_m)$ be some permutation of $\{0,1,\ldots,m\}$. Assume that all solution components $u_j$ are the same functions, which differ only by time shifts, and the order of shifts is determined by $\pi$:

$$u_j(t) = u(t + k_j \Delta). \tag{7}$$

Let $T(\Delta)$ be the period of $u(t)$. Make the substitution (7) in (6); after time normalization $t + k_j\Delta \mapsto t$, we obtain:

$$\dot{u} = -\beta u + \alpha\left(u(t-1) + \sum_{s=0, s\neq j}^{m} u\left(t + (k_s - k_j)\Delta\right)\right) F\left(u(t-1) + \sum_{s=0, s\neq j}^{m} u\left(t + (k_s - k_j)\Delta\right)\right). \quad (8)$$

Suppose that

$$pT(\Delta) = (m+1)\Delta. \quad (9)$$

for some $p \in \mathbb{N}$. Then, regardless of $\pi$,

$$\sum_{s=0, s\neq j}^{m} u\left(t + (k_s - k_j)\Delta\right) = \sum_{s=1}^{m} u\left(t - s\Delta\right).$$

Thus, we shall find a $T$-periodical solution of the equation

$$\dot{u} = -\beta u + \alpha\left(u(t-1) + \sum_{s=1}^{m} u(t - s\Delta)\right) F\left(u(t-1) + \sum_{s=1}^{m} u(t - s\Delta)\right) \quad (10)$$

and $\Delta > 0$, such that the condition (9) holds for some $p \in \mathbb{N}$.

## 3. ANALYSIS OF THE AUXILIARY EQUATION

### 3.1 Time delays

The relay equation (10) has delays $\{1, \Delta, \ldots, m\Delta\}$; arrange them in increasing order:

$$0 < \tau_0 < \tau_1 < \cdots < \tau_m.$$

Now the equation (10) takes the form

$$\dot{u} = -\beta u + \alpha\left(\sum_{s=0}^{m} u(t - \tau_s)\right) F\left(\sum_{s=0}^{m} u(t - \tau_s)\right). \quad (11)$$

After substitutions

$$t/\tau_0 \mapsto t, \ u(\tau_0 t) \mapsto u(t), \ \alpha\tau_0 \mapsto \alpha, \ \beta\tau_0 \mapsto \beta, \ \tau_k/\tau_0 \mapsto \tau_k \ (k = 0, 1, \ldots, m), \quad (12)$$

the equation (11) does not change, and $\tau_0 = 1$. Then the set of delays takes the form

$$\{1, \tau_1, \tau_2, \ldots, \tau_m\} = \begin{cases} \{1, 1/\Delta, 2, \ldots, m\}, & \text{if } \Delta \leq 1, \\ \{1, \Delta, 2\Delta, \ldots, m\Delta\}, & \text{if } \Delta \geq 1. \end{cases}$$

The smallest delay of the equation (11) will be 1, and

$$\tau_m = \begin{cases} \max\{1/\Delta, m\}, & \text{if } \Delta \leq 1, \\ m\Delta, & \text{if } \Delta \geq 1, \end{cases}$$

is the largest delay.

Denote

$$w(t) = \sum_{k=0}^{m} u(t - \tau_k), \tag{13}$$

so the equation (11) takes a more compact form

$$\dot{u} = -\beta u + \alpha w(t) F(w(t)). \tag{14}$$

### 3.2 The set of initial functions

Let $u_0 > 1$. We introduce the initial set (Fig. 3) of continuous functions on $[-\tau_m; 0]$:

$$S = \{\varphi \in C[-\tau_m, 0]: \varphi(t) > 1 \text{ for } t \in [-\tau_m, 0), \varphi(0) = u_0\}. \tag{15}$$

### 3.3 The switching points

We will say that a point $t$ is a *switching point* if the right side of equation (14) changes an analytical form in $t$. Switching points may be one of two kinds:

1. The roots of the equation $w(t) = 1$. Denote them by

$$t_k, \quad k = 0, 1, \ldots \tag{16}$$

   These are discontinuities of the right side of the equation (14); the function $F(w(t))$ "jumps" at these points from 0 to 1, or vice versa.

2. Switching points of kind (A), increased by delays $\tau_0 = 1, \tau_1, \ldots, \tau_m$, at which $w(t) < 1$. At these points $t$, $F(w(t)) = 1$, then the right side of the equation (14) takes a form

$$-\beta u + \alpha(u(t - \tau_0) + u(t - \tau_1) + \cdots + u(t - \tau_m)),$$

   and one of terms $u(t - \tau_i)$ ($i = 0, 1, \ldots, m$) changes analytic form at point $t$.

We are looking for a periodic solution *having a minimal number of switching points*.

**Proposition 3.1.** *Periodic solutions of the equation (14) have at least two switching points $t_0, t_1$ of the kind (A).*

*Proof.* The function $w(t)$ is periodic (as a sum of periodic functions), thus the number of switching points of kind (A) is even.

Suppose this number is zero. Then there are no switching points of the kind (B) too, and the equation (14) has the same form for $t > 0$. Let $\varphi \in S$, then $\varphi(t) > 1$ for $[-\tau_m, 0]$, therefore $F(w(t))$ is zero for $t \in [0, \tau_m]$, and, by our assumption, for $t > 0$. Thus $u(t)$ is a solution of the Cauchy problem

$$\dot{u} = -\beta u, \quad u|_{t=0} = u_0.$$

So, $u_\varphi(t) = u_0 e^{-\beta t}$ for $t > 0$, but this solution is not periodic.

Therefore, the number of switching points is at least two.

□

*3.4 Solution of the auxiliary equation*

Introduce the following notation:

$$A = \sum_{i=0}^{m} e^{\beta \tau_i} = e^\beta + e^{\beta \tau_1} + \cdots + e^{\beta \tau_m},$$

$$\tau_* = \min\{2, \tau_1\} = \begin{cases} \min\{2, 1/\Delta\}, & \text{if } \Delta < 1, \\ \min\{2, \Delta\}, & \text{if } \Delta > 1, \end{cases}$$

$$s_1 = t_1 - t_0.$$

**Theorem 3.2**. *Let the parameters $\alpha, \beta$ and $\tau_*$ hold the following conditions:*

| | |
|---|---|
| $\dfrac{\alpha}{\beta} e^\beta \left( \ln\left(\dfrac{\beta}{\alpha}\right) + 1 \right) \geq 1,$ | (17) |
| $e^{\beta \tau_*} - 1 \leq \alpha e^\beta (\tau_* - 1)$ or $\begin{cases} e^{\beta \tau_*} - 1 > \alpha e^\beta (\tau_* - 1), \\ \dfrac{1}{\beta} \ln \dfrac{\alpha}{\beta} < \tau_* - 1, \end{cases}$ | (18) |
| $e^{-\beta s_1}(e^{-\beta} + \alpha s_1) > 1$ | (19) |
| $\dfrac{\alpha^2}{2} e^\beta (s_1 - 1)^2 + \alpha s_1 > \dfrac{e^{\beta \tau_k} - 1}{\sum_{i=0}^{k-1} e^{\beta \tau_i}},$ if $s_1 \leq \tau_k - \tau_{k-1}, \quad k = 1, \ldots, m,$ | (20) |
| $\dfrac{\alpha^2}{2} e^\beta (s_1 - 1)^2 + \alpha s_1 > \dfrac{e^{\beta(s_1 + \tau_{k-1})} - 1}{\sum_{i=0}^{k-1} e^{\beta \tau_i}},$ if $s_1 > \tau_k - \tau_{k-1}, \quad k = 1, \ldots, m,$ | (21) |
| $\dfrac{\alpha^2}{2} e^\beta (s_1 - 1)^2 + \alpha s_1 > \dfrac{(e^{\beta s_1} - \alpha s_1) e^{\beta \tau_k} - 1}{\sum_{i=0}^{k-1} e^{\beta \tau_i}}, \quad k = 1, \ldots, m,$ | (22) |

*Then the equation (14) has a periodic solution $u_*$, which is independent of the initial function $\varphi \in S$. Let $T$ be the period of $u_*$. Then the following statements are true.*

1. *The function $u_*(t)$ has a minimal number of switching points on a period: $t_0$, $t_0 + 1$, $t_1$.*
   $t_0 = \frac{1}{\beta}\ln(u_0 A)$, $\quad 1 \leq t_1 - t_0 \leq \tau_*$, $\quad e^{\beta(t_1 - t_0)} - 1 = \alpha e^\beta (t_1 - t_0 - 1)$.

2. *The function $u_*(t)$ and its period $T$ have the following analytical form:*

$$u_*(t) = \begin{cases} u_0 e^{-\beta t}(\alpha A(t - t_0) + 1) & \text{if } t \in [t_0, t_0 + 1]. \\ u_0 e^{-\beta t}\left(\frac{\alpha^2}{2} A e^\beta (t - t_0 - 1)^2 + \alpha A(t - t_0) + 1\right) & \text{if } t \in [t_0 + 1, t_1], \\ u_0 e^{-\beta t}\left(\frac{\alpha^2}{2} A e^\beta (t_1 - t_0 - 1)^2 + \alpha A(t_1 - t_0) + 1\right) & \text{if } t \in [t_1, t_0 + T], \end{cases} \quad (23)$$

$$u_*(t + T) \equiv u_*(t),$$

$$T = \frac{1}{\beta}\ln\left(\frac{\alpha^2}{2} A e^\beta (t_1 - t_0 - 1)^2 + \alpha A(t_1 - t_0) + 1\right). \quad (24)$$

A schematic plot of function $u_*$ is shown on Fig. 4.

Now we will prove this theorem.

### 3.5 Construction of the solution

We now construct a solution of the equation (14) by considering consequently the intervals between the switching points.

#### 3.5.1. Step 1

We start with the interval $[0; \tau_m]$. $w(t) > 1$ on that interval, because $w(t) \geq u(t - \tau_m) = \varphi(t - \tau_m) > 1$ for each initial function $\varphi \in S$.

Thus $F(w(t)) = 0$, and $u(t)$ is the solution of a Cauchy problem:

$$\dot{u} = -\beta u, \quad u|_{t=0} = u_0, \quad (25)$$

i. e. $u(t) = u_0 e^{-\beta t}$.

For $t > \tau_m$,

$$w(t) = u_0 e^{-\beta(t-1)} + u_0 e^{-\beta(t-\tau_1)} + \cdots + u_0 e^{-\beta(t-\tau_m)} = u_0 A e^{-\beta t}. \quad (26)$$

The first switching point $t_0$ is a root of the equation $w(t) = 1$,

$$t_0 = \frac{1}{\beta}\ln(u_0 A). \tag{27}$$

The solution preserves its analytical form until $w(t) > 1$. Therefore

$$u(t) = u_*(t) = u_0 e^{-\beta t} \text{ for } t \in [0, t_0]. \tag{28}$$

### 3.5.2. Construction of solution for $w(t) < 1$

As $t > t_0$, $F(w(t))$ becomes nonzero. Now we need to prove the following lemma.

**Lemma 3.3**. *For each step, until $w(t) < 1$, $u(t)$ can be found as a solution of Cauchy problem:*

$$\dot{u} = -\beta u + \alpha u_0 e^{-\beta t}\tilde{p}(t), \quad u|_{t=\tilde{t}} = u_0 e^{-\beta \tilde{t}}\tilde{u}, \tag{29}$$

where $\tilde{t}, \tilde{u}$ and polynome $\tilde{p}(t)$ are uniquely defined from the previous step. The solution of (29) has the form

$$u(t) = u_0 e^{-\beta t}\left(\tilde{u} + \alpha \int_{\tilde{t}}^{t} \tilde{w}(s)ds\right). \tag{30}$$

*Proof.* We prove it by induction. We already proved the base case (step 1): see eq. (25), here $\tilde{p}(t) \equiv 0$, $\tilde{t} = 0$, $\tilde{u} = 1$ and its solution is eq. (28).

Now we prove the induction step. Suppose that for each step $1, \dots, k$ solution has the form (30), and on a step $k$ we have

$$u(t) = u_0 e^{-\beta t} p(t)$$

for some polynomial $p(t)$. Then

$$w(t) = u(t-1) + u(t-\tau_1) + \dots + u(t-\tau_m)$$

has the same form: $u(t) = u_0 e^{-\beta t} q(t)$ for some polynomial $q(t)$.

Therefore $u(t)$ on step $k+1$ can be found as a solution of Cauchy problem (29), $\tilde{p}(t) = q(t)$, $\tilde{t}$ — the end of the time interval at step $k$, $\tilde{u} = p(\tilde{t})$. Implicit integration of (29) gives (30), which ends the proof.

□

### 3.5.3. Step 2

The next interval is $[t_0, \min\{t_0 + 1, t_1\}]$. We will show that $t_0 + 1 < t_1$, i. e. the switching point $t_1$ lies outside the interval $[t_0, t_0 + 1]$.

**Lemma 3.4.** $w(t) < 1$ for $t \in (t_0, t_0 + 1]$.

*Proof.* For $t \in (t_0, t_0 + 1]$, the values $t - 1, t - \tau_1, \ldots, t - \tau_m$ belong to $(0; t_0]$, so $u(t)$ has the form (28).

Then $w(t)$ has the form (26), and $w(t) = 1$ is equivalent to

$$u_0 A e^{-\beta t} = 1.$$

This equation has only one root (27), hence $w(t) < 1$ for $t \in (t_0, t_0 + 1]$

□

Therefore, the next interval is $[t_0, t_0 + 1]$.

Solving (29) with $\tilde{p}(t) \equiv A$, $\tilde{t} = t_0$, $\tilde{u} = 1$, we obtain

$$u(t) = u_*(t) = u_0 e^{-\beta t}(\alpha A(t - t_0) + 1) \text{ for } t \in [t_0, t_0 + 1]. \tag{31}$$

### 3.5.4. Step 3

The next interval is $[t_0 + 1, \min\{t_0 + \tau_*, t_1\}]$, where $\tau_* = \min\{2, \tau_1\}$. Let us show that $t_1 \in [t_0 + 1, t_0 + \tau_*]$. On this interval,

$$w(t) = u_0 e^{-\beta(t-1)}(\alpha A(t - t_0 - 1) + 1) + u_0 e^{-\beta(t-\tau_1)} + \cdots + u_0 e^{-\beta(t-\tau_m)} =$$
$$= u_0 A e^{-\beta t}\left(1 + \alpha e^\beta (t - t_0 - 1)\right). \tag{32}$$

After substitutions $u_0 A = e^{\beta t_0}$, $s := t - t_0$, $s \in [1, \tau_*]$, equation $w(t) = 1$ takes the form:

$$e^{\beta s} - 1 = \alpha e^\beta (s - 1). \tag{33}$$

Let

$$f(s) = e^{\beta s} - 1, \quad g(s) = \alpha e^\beta (s - 1).$$

$f(s)$ is increasing, convex, and $f(0) = 0$; $g(s)$ is affine, and $g(1) = 0$.

Now we find the conditions that (i) a root $s > 1$ of eq. (33) exists; (ii) $s \in (1, \tau_*]$.

**Lemma 3.5.** *A root $s_1 > 1$ of eq. (33) exists iff condition (17) holds.*

*Proof.* Find the minimum $s_0$ of function $f(s) - g(s)$:

$$(f(s) - g(s))' = 0 \iff \beta e^{\beta s} - \alpha e^\beta = 0 \iff s_0 = 1 + \frac{1}{\beta} \ln \frac{\alpha}{\beta}.$$

A direct check shows that condition (11) is equivalent to $f(s_0) - g(s_0) \leq 0$, so by continuity, a root $s_1$ exists.

□

**Lemma 3.6.** *Assume that the condition (11) holds. The (lesser) root $s_1$ of eq. (33) belongs to $[1, \tau_*]$ iff one of the following conditions holds:*

1. $e^{\beta \tau_*} - 1 \leq \alpha e^{\beta}(\tau_* - 1)$,

2. $e^{\beta \tau_*} - 1 > \alpha e^{\beta}(\tau_* - 1)$ and $\frac{1}{\beta} \ln \frac{\alpha}{\beta} \leq \tau_* - 1$.

*Proof.* Note that $f(1) - g(1) = e^{\beta} - 1 > 0$. Then a zero of function $(f(s) - g(s))$ belongs to $[1; \tau_*]$ iff one of the following conditions holds (see fig. 5):

1. $f(\tau_*) \leq g(\tau_*)$,

2. $f(\tau_*) > g(\tau_*)$, and $s_0 \in [1, \tau_*]$.

After substitutions for $f(\tau_*)$, $g(\tau_*)$ and $s_0$, we obtain the reqired.

□

As a result, we proved that if the conditions (17) and (18) hold, then eq. (14) has a root $s_1 \in [1; \tau_*)$. Thus the root $t_1 = t_0 + s_1$ of eq. $w(t) = 1$ also exists and belongs to $[t_0 + 1, t_0 + \tau_*]$. Then the next interval of solution $u(t)$ construction is $[t_0 + 1, t_1]$. By eq. (32) and Lemma 3 for $\tilde{p}(t) \equiv A\left(1 + \alpha e^{\beta}(t - t_0 - 1)\right)$, $\tilde{t} = t_0 + 1$, $\tilde{u} = \alpha A + 1$, we obtain

$$u(t) = u_*(t) = u_0 e^{-\beta t}\left(\frac{\alpha^2}{2} A e^{\beta}(t - t_0 - 1)^2 + \alpha A(t - t_0) + 1\right) \text{ for } t \in [t_0 + 1, t_1]. \quad (34)$$

### 3.5.5. Parameters $\alpha, \beta, \tau_*$ domain, over which $t_1 \in [t_0 + 1, t_0 + \tau_*]$

The conditions (17) and (18) *implicitly* define the domain of the parameters $\alpha, \beta, \tau_*$, over which $t_1 \in [t_0 + 1, t_0 + \tau_*]$. Now we define it explicitly.

Express $\alpha$ from conditions (17) and (18). Transform eq. (17):

$$\frac{\alpha}{\beta} e^{\beta}\left(\ln \frac{\beta}{\alpha} + 1\right) \leq 1 \iff 1 + \ln \frac{\beta}{\alpha} \leq \frac{\beta e^{-\beta}}{\alpha}.$$

Apply the exponent:

$$\frac{e\beta}{\alpha} \leq \exp \frac{\beta e^{-\beta}}{\alpha} \iff \frac{-\beta e^{-\beta}}{\alpha} \exp \frac{-\beta e^{-\beta}}{\alpha} \geq -e^{-\beta-1} \iff \frac{-\beta e^{-\beta}}{\alpha} \geq W(-e^{-\beta-1}),$$

where $W$ is the Lambert W-function, i. e. inverse to the function $x \mapsto x e^x$ defined for $x \in [-1; +\infty)$. Therefore, we may rewrite condition (17) in the explicit form:

$$\alpha \geq -\frac{\beta e^{-\beta}}{W(-e^{-\beta-1})} \stackrel{\text{def}}{=} \varphi(\beta). \quad (45)$$

Inequalities from (18) may be rewritten as:

$$e^{\beta \tau_*} - 1 \leq \alpha e^\beta (\tau_* - 1) \Leftrightarrow \alpha \geq \frac{e^{\beta \tau_*} - 1}{e^\beta (\tau_* - 1)} \stackrel{\text{def}}{=} \theta(\beta, \tau_*). \tag{36}$$

$$\frac{1}{\beta} \ln \frac{\alpha}{\beta} \leq \tau_* - 1 \Leftrightarrow \alpha \leq \beta e^{\beta(\tau_* - 1)} \stackrel{\text{def}}{=} \delta(\beta, \tau_*). \tag{37}$$

Now we combine these conditions into one.

**Lemma 3.7.** *The root $t_1$ of the equation $w(t) = 1$ (found at the Step 3) belongs to the interval $[t_0 + 1, t_0 + \tau_*]$ iff the following restrictions hold.*

$$\begin{cases} \alpha \geq \theta(\beta; \tau_*) & \text{for } \beta < \tilde{\beta}(\tau_*); \\ \alpha \geq \varphi(\beta) & \text{for } \beta \geq \tilde{\beta}(\tau_*), \end{cases} \tag{38}$$

$$\tilde{\beta}(\tau_*) = \frac{1}{\tau_* - 1} + \frac{1}{\tau_*} W\left( \frac{\tau_*}{1 - \tau_*} e^{\frac{\tau_*}{1-\tau_*}} \right),$$

$\varphi(\beta)$ *defined by eq. (35), and $\theta(\beta; \tau_*)$ defined by eq. (36).*

*Proof.* In the following propositions, we describe the relative positions of curves $\varphi(\beta)$, $\theta(\beta, \tau_*)$ and $\delta(\beta, \tau_*)$.

**Proposition 3.8.** $\varphi(\beta) \geq \delta(\beta, \tau_*) \Leftrightarrow 0 \leq \beta \leq \tilde{\beta}(\tau_*)$, *and the equality holds only for $\beta = 0$ and $\beta = \tilde{\beta}(\tau_*)$.*

The equality for $\beta = 0$ is clear, consider $\beta > 0$.

$$-\frac{\beta e^{-\beta}}{W(-e^{-\beta-1})} \geq \beta e^{\beta(\tau_* - 1)} \Leftrightarrow W(-e^{-\beta-1}) \geq -e^{-\beta \tau_*} \Leftrightarrow \ldots$$

Apply the transformation $x \mapsto xe^x$ to the inequality (this is an increasing bijection for $x \geq -1$).

$$\begin{aligned} \ldots &\Leftrightarrow -e^{-\beta-1} \geq -e^{-\beta \tau_*} \exp(-e^{-\beta \tau_*}) \Leftrightarrow e^{-\beta-1+\beta \tau_*} \leq \exp(-e^{-\beta \tau_*}) \Leftrightarrow \\ &\Leftrightarrow -\beta - 1 + \beta \tau_* \leq -e^{-\beta \tau_*} \Leftrightarrow e^{\beta \tau_*}(\beta(\tau_* - 1) - 1) \leq -1 \Leftrightarrow \\ &\Leftrightarrow \left( \beta \tau_* + \frac{\tau_*}{1 - \tau_*} \right) e^{\beta \tau_* + \frac{\tau_*}{1-\tau_*}} \leq \frac{\tau_*}{1 - \tau_*} e^{\frac{\tau_*}{1-\tau_*}}. \end{aligned} \tag{39}$$

The function $f(x) = xe^x$ takes the value $\frac{\tau_*}{1-\tau_*} e^{\frac{\tau_*}{1-\tau_*}}$ at exactly two points: $x = \frac{\tau_*}{1-\tau_*} < -1$ and $x = W\left( \frac{\tau_*}{1-\tau_*} e^{\frac{\tau_*}{1-\tau_*}} \right) \in (-1; 0)$. Thus, the preceding inequality is equivalent to the following (see Fig. 6):

$$\frac{\tau_*}{1 - \tau_*} \leq \beta \tau_* + \frac{\tau_*}{1 - \tau_*} \leq W\left( \frac{\tau_*}{1 - \tau_*} e^{\frac{\tau_*}{1-\tau_*}} \right). \tag{40}$$

The left inequality is equivalent to $\beta \geq 0$, and the right side is equivalent to $\beta \leq \tilde{\beta}(\tau_*)$.

□

Note that we can turn all inequalities in this proof into equalities. It follows that $\beta = \tilde{\beta}(\tau_*)$ is only positive root of the equation

$$e^{\beta \tau_*}(\beta(\tau_* - 1) - 1) = -1. \tag{41}$$

**Proposition 3.9.** $\delta(\beta, \tau_*) = \theta(\beta, \tau_*) \Leftrightarrow \beta = 0$ or $\beta = \tilde{\beta}(\tau_*)$.

$$\beta e^{\beta(\tau_*-1)} = \frac{e^{\beta \tau_*} - 1}{e^{\beta}(\tau_* - 1)} \Leftrightarrow e^{\beta \tau_*}(\beta(\tau_* - 1) - 1) = -1.$$

This equation is the same as eq. (41), which, as already proven, has roots $\beta = 0$ and $\beta = \tilde{\beta}(\tau_*)$.

□

**Proposition 3.10.** $\theta(\beta, \tau_*) \geq \varphi(\beta)$, and the equality holds only for $\beta = 0$ and $\beta = \tilde{\beta}(\tau_*)$.

Fix $\beta$, find the minimum $\tau_*^{min}$ of the function $\theta(\beta, \tau_*)$.

$$\frac{d}{d\tau_*} \theta(\beta, \tau_*) = \frac{\beta e^{\beta \tau_*}(\tau_* - 1) - (e^{\beta \tau_*} - 1)}{e^{\beta}(\tau_* - 1)^2},$$

$$\frac{d}{d\tau_*} \theta(\beta, \tau_*) = 0 \Leftrightarrow e^{\beta \tau_*}(\beta(\tau_* - 1) - 1) = -1.$$

This condition coincides with (41), but now we need to solve it with respect to $\tau_*$.

$$e^{\beta \tau_*}(\beta(\tau_* - 1) - 1) = -1 \Leftrightarrow e^{\beta(\tau_*-1)-1}(\beta(\tau_* - 1) - 1) = -e^{-\beta-1} \Leftrightarrow$$

$$\Leftrightarrow \beta(\tau_* - 1) - 1 = W(-e^{-\beta-1}) \Leftrightarrow \tau_* = \frac{W(-e^{-\beta-1}) + 1}{\beta} + 1.$$

This point is the minimum, because

$$\lim_{\tau_* \to 1+0} \theta(\beta, \tau_*) = \lim_{\tau_* \to +\infty} \theta(\beta, \tau_*) = +\infty.$$

□

Note that if we initially fix $\tau_*$, then the function $\theta(\beta, \tau_*)$ also has exactly one "touch point" with $\varphi(\beta)$: this is the point $(\tilde{\beta}(\tau_*), \tau_*)$, and for $\beta \neq \tilde{\beta}(\tau_*)$, by Proposition 3.10, a strict inequality $\theta(\beta, \tau_*) > \varphi(\beta)$ holds.

**Proposition 3.11.** $\delta(\beta, \tau_*) \geq \theta(\beta, \tau_*)$ for $\beta \geq \tilde{\beta}(\tau_*)$.

By Proposition 3.9, $\delta(\beta, \tau_*)$ and $\theta(\beta, \tau_*)$ have no intersections for $\beta > \tilde{\beta}(\tau_*)$.

$$\theta(\beta, \tau_*) \sim e^{\beta \tau_*} \text{ as } \beta \to +\infty, \quad \delta(\beta, \tau_*) = \beta e^{\beta \tau_*},$$

so, the function $\delta(\beta, \tau_*)$ increases faster. By continuity, $\delta(\beta, \tau_*) > \theta(\beta, \tau_*)$ for $\beta \geq \tilde{\beta}(\tau_*)$.

□

Thus Lemma 3.7 follows from Propositions 3.9, 3.10 and 3.11. On Fig. 7, see the examples of parameters $\alpha$ and $\beta$ domains with $\tau_*$ fixed.

### 3.5.6 Step 4

The next interval is $[t_1, t_2]$. Here $w(t) = 1$, $F(w(t)) = 0$, so the solution $u(t)$ can be found from the Cauchy problem:

$$\dot{u} = -\beta u, \quad u|_{t=t_1} = u_1, \quad u_1 \stackrel{\text{def}}{=} u(t_1) = u_0 e^{-\beta t_1} \left( \frac{\alpha^2}{2} A e^\beta (s_1 - 1)^2 + \alpha A s_1 + 1 \right).$$

Therefore

$$u(t) = u_*(t) = u_1 e^{-\beta(t-t_1)} = u_0 e^{-\beta t} \left( \frac{\alpha^2}{2} A e^\beta (s_1)^2 + \alpha A s_1 + 1 \right) \text{ for } t \in [t_1, t_2]. \tag{42}$$

### 3.6. Periodicity of the solution $u(t)$

We have already constructed the solution of eq. (14) on the interval $[0, t_2]$ with the initial function from the set (15). The solution on this interval coincides with the function $u_*(t)$, which is defined by eq. (28), (31), (34), (42). If $t_2 - t_1 > \tau_m$, then the solution may be periodically extended with the period $T = t_2 - t_0$. Now we consider the function $w(t)$ over the interval $[t_1, t_1 + \tau_m]$, and prove that $w(t) > 1$.

### 3.6.1 Behavior of the function $w(t)$

The analytical form of the function $w(t)$ changes at the switching points of $u_*(t)$, increased by delays $\tau_k$, i. e. at the points from the following sets:

$$E_* \stackrel{\text{def}}{=} \{t_0 + \tau_k\}, \quad E_{**} \stackrel{\text{def}}{=} \{t_0 + 1 + \tau_k\}, \quad E_{***} \stackrel{\text{def}}{=} \{t_1 + \tau_k\}, \text{ for } k = 0, 1, \ldots, m, \quad \tau_0 = 1.$$

Note that these sets are not necessarily disjoint.

Define the functions

$$\xi(s) \stackrel{\text{def}}{=} \alpha e^{-\beta s} s, \quad \eta(s) \stackrel{\text{def}}{=} \frac{\alpha^2}{2} e^{-\beta(s-1)} (s-1)^2$$

**Lemma 3.12**. *The function $w(t)$ changes its analytical form at the points from $E_*$, $E_{**}$, $E_{***}$ by the following formulas:*

| | |
|---|---|
| $w(t_0 + \tau_k + 0) = w(t_0 + \tau_k - 0) + \xi(t - \tau_k - t_0),$ | (43) |
| $w(t_0 + 1 + \tau_k + 0) = w(t_0 + 1 + \tau_k - 0) + \eta(t - \tau_k - t_0),$ | (44) |

$$w(t_1 + \tau_k + 0) = w(t_1 + \tau_k - 0) - \xi(t - \tau_k - t_0) - \eta(t - \tau_k - t_0) + (u_1 e^{\beta t_1} - u_0)e^{-\beta(t-\tau_k)}. \quad (45)$$

*If a point belongs to more than one set, then the correspondent addition terms are summed up.*

*Proof.* Consider a point $t_0 + \tau_k \in E_*$. At this point the term $u(t - \tau_k)$ changes its form from (28) to (31). Then the function $w(t)$ changes by

$$u_0 e^{-\beta(t-\tau_k)}(\alpha A(t - \tau_k - t_0) + 1) - u_0 e^{-\beta(t-\tau_k)} = \alpha e^{-\beta(t-\tau_k-t_0)}(t - \tau_k - t_0).$$

at $t_0 + \tau_k$.

Formulas (44) and (45) are proved the same way: the analytic form of one of the terms changes from (31) to (34), if $t \in E_{**}$, and from (34) to (42), if $t \in E_{***}$.

Even if some of points in $E_*, E_{**}, E_{***}$ coincide, they affect to different terms of (13), so these additions are independent.

□

Note that the additions in (43), (44) are non-negative, and take zero values at the correspondent switching points.

**Lemma 3.13.** *Let the conditions (17) and (18) hold, and, moreover, the parameters satisfy conditions (19) – (22). Then $w(t) > 1$ for $t \in (t_1, t_1 + \tau_m]$.*

*3.6.2 Proof of Lemma 3.13*

The idea of the proof is as follows: for each interval $(t_1, t_1 + \tau_m]$ we define a function $\tilde{v}(t) < w(t)$ and show that $1 < \tilde{v}(t)$. From this, $w(t) > 1$ follows.

We will successively consider the intervals $(t_1, t_1 + 1]$, $(t_1 + \tau_{k-1}, t_1 + \tau_k]$ for $k = 1, \dots, m$. On each step, we prove that $w(t) > 1$ over the considered interval, therefore it does not contain any new switching points. It follows that the function $w(t)$ changes its analytical form *only* at switching points listed in Lemma 3.12.

Let $\tilde{u}(t)$ be the function defined on the interval $[0, t_2]$ (see fig. 8):

$$\tilde{u}(t) \stackrel{\text{def}}{=} \begin{cases} u_0 e^{-\beta t} & \text{for } t \in [0, t_1], \\ u_1 e^{-\beta(t-t_1)} & \text{for } t \in (t_1, t_1 + 1 + \tau_m]. \end{cases} \quad (46)$$

Define the function $\tilde{w}(t) \stackrel{\text{def}}{=} \sum_{i=0}^{m} \tilde{u}(t - \tau_i)$ on the interval $[\tau_m, t_1 + \tau_m]$:

$$\tilde{w}(t) = \begin{cases} e^{-\beta(t-t_0)} & \text{for } t \in [\tau_m, t_1 + 1], \\ C_k e^{-\beta(t-t_0)} & \text{for } t \in (t_1 + \tau_{k-1}, t_1 + \tau_k], \ k = 1, \dots, m, \end{cases} \quad (47)$$

where

$$C_k = 1 + \left(\frac{\alpha^2}{2} e^\beta (s_1 - 1)^2 + \alpha s_1\right) \sum_{i=0}^{k-1} e^{\beta \tau_i}.$$

The function $\widetilde{w}(t)$ has jump discontinuities at the points $t_1 + \tau_k \in E_{***}$:

$$\widetilde{w}(t_1 + \tau_k + 0) = \widetilde{w}(t_1 + \tau_k - 0) + e^{-\beta(t-\tau_k)}(u_1 e^{\beta t_1} - u_0). \tag{48}$$

Note that $\widetilde{w}(t)$ changes its form only at these points (see Fig. 9).

For each interval $[t_0 + 1, t_1 + 1]$, $[t_0 + \tau_k, t_1 + \tau_k] \cap [t_1 + \tau_{k-1}, t_1 + \tau_k]$, $k = 1, \ldots, m$, we apply the correspondent addition from (43) to function $\widetilde{w}(t)$.

Let

$$\tilde{\xi}(s) = \begin{cases} \xi(s) & \text{for } s \in [0, s_1], \\ 0 & \text{for } s \notin [0, s_1]. \end{cases}$$

Define a function $\tilde{v}(t)$ over interval $[\tau_m, t_1 + \tau_m]$ (see Fig. 9):

$$\tilde{v}(t) \stackrel{\text{def}}{=} \begin{cases} \widetilde{w}(t) + \tilde{\xi}(t - t_0 - 1) & \text{for } t \in [\tau_m, t_1 + 1], \\ \widetilde{w}(t) + \tilde{\xi}(t - t_0 - \tau_k) & \text{for } t \in (t_1 + \tau_{k-1}, t_1 + \tau_k], \ k = 1, \ldots, m. \end{cases}$$

**Proposition 3.14.** $\widetilde{w}(t) \le \tilde{v}(t) \le w(t)$ if $t \in [\tau_m, t_1 + \tau_m]$.

*Proof.* By definition, functions $\widetilde{w}(t)$ and $\tilde{v}(t)$ differ by additions (43); functions $\tilde{v}(t)$ and $w(t)$ differ at least by additions (44).

$\square$

**Proposition 3.15.** *Let the conditions (17) – (22) hold. Then $\tilde{v}(t) > 1$ for $t \in (t_1, t_1 + \tau_m]$.*

*Proof.* Express $\tilde{v}(t)$ explicitly.

$$\tilde{v}(t) = \begin{cases} e^{-\beta(t-t_0)} & \text{for } t \in [\tau_m, t_0 + 1], \\ e^{-\beta(t-t_0)} + \alpha e^{-\beta(t-t_0-1)}(t - t_0 - 1) & \text{for } t \in (t_0 + 1, t_1 + 1]. \end{cases}$$

Depending on order of points $t_0 + \tau_k$ and $t_1 + \tau_{k-1}$, $\tilde{v}(t)$ takes following forms over $[t_1 + \tau_{k-1}, t_1 + \tau_k]$:

1. if $t_0 + \tau_k \in [t_1 + \tau_{k-1}, t_1 + \tau_k]$, then

$$\tilde{v}(t) = \begin{cases} C_k e^{-\beta(t-t_0)} & \text{for } t \in (t_1 + \tau_{k-1}, t_0 + \tau_k], \\ C_k e^{-\beta(t-t_0)} + \alpha e^{-\beta(t-t_0-\tau_k)}(t - t_0 - \tau_k) & \text{for } t \in (t_0 + \tau_k, t_1 + \tau_k]; \end{cases} \tag{49}$$

2. if $t_0 + \tau_k < t_1 + \tau_{k-1}$, then

$$\tilde{v}(t) = C_k e^{-\beta(t-t_0)} + \alpha e^{-\beta(t-t_0-\tau_k)}(t - t_0 - \tau_k) \text{ for } t \in (t_1 + \tau_{k-1}, t_1 + \tau_k]. \tag{50}$$

The function $C_k e^{-\beta s}$ is decreasing by $s$. The function $C_k e^{-\beta \tau_k} e^{-\beta s} + \alpha e^{-\beta s} s$ has local maximum at point $s_{\max} = \frac{\alpha - \beta C_k e^{-\beta \tau_k}}{\alpha \beta}$. So, over arbitrary interval it is either monotonic, or has a local maximum: in both cases we need check inequalities only at bounds of the interval. It follows that the sufficient conditions for $\tilde{v}(t) > 1$ over $t \in (t_1, t_1 + 1]$, and $t \in (t_1 + \tau_{k-1}, t_1 + \tau_k]$, $k = 1, \ldots, m$, are:

$$e^{-\beta(t-t_0)} + \alpha e^{-\beta(t-t_0-1)}(t - t_0 - 1)|_{t=t_1} \geq 1, \tag{51}$$

$$e^{-\beta(t-t_0)} + \alpha e^{-\beta(t-t_0-1)}(t - t_0 - 1)|_{t=t_1+1} > 1, \tag{52}$$

$$C_k e^{-\beta(t-t_0)} + \alpha e^{-\beta(t-t_0-\tau_k)}(t - t_0 - \tau_k)|_{t=\max\{t_0+\tau_k, t_1+\tau_{k-1}\}} > 1, \tag{53}$$

$$C_k e^{-\beta(t-t_0)} + \alpha e^{-\beta(t-t_0-\tau_k)}(t - t_0 - \tau_k)|_{t=t_1+\tau_k} > 1. \tag{54}$$

The inequality (51) is true, because $s_1 = t_1 - t_0$ is a root of the equation $e^{\beta s} - 1 = \alpha e^{\beta}(s-1)$ by definition. The inequality (52) is equivalent to the condition (19). The inequality (53) for $t_0 + \tau_k \geq t_1 + \tau_{k-1}$ turns into (20). For $t_0 + \tau_k < t_1 + \tau_{k-1}$, a stronger inequality

$$C_k e^{-\beta(t-t_0)}|_{t=t_1+\tau_{k-1}} > 1$$

equivalent to the condition (21). The inequality (54) is equivalent to the condition (22).

□

It follows from Propositions 3.14 and 3.15, that Lemma 3.13 is true.

## 4. DISCRETE TRAVELLING WAVES

In this section, we will prove that for fixed $\alpha, \beta$ and for some $\Delta$ the system (9) has a periodical solution with period $T(\Delta)$, such that the equation (9) holds for $p = 1$.

If $\Delta > 1$, then a set of time delays is $\{1, \Delta, 2\Delta, \ldots, m\Delta\}$. In that case, substitutions (12) do not affect the equation (9). Substitute $T(\Delta)$ from (24) to (9) for $p = 1$:

$$\frac{1}{\beta} \ln\left(\frac{\alpha^2}{2} A e^{\beta}(s_1 - 1)^2 + \alpha A s_1 + 1\right) = \Delta(m + 1). \tag{55}$$

After simple transformations we obtain:

$$\frac{1}{A}\left(e^{\beta \Delta(m+1)} - 1\right) = \frac{\alpha^2}{2} e^{\beta}(s_1 - 1)^2 + \alpha s_1. \tag{56}$$

**Proposition 4.1.** $\lim\limits_{\alpha \to +\infty} s_1(\alpha) = 1 + 0$.

*Proof.* By definition, $s_1$ is the minimal root of the equation $\alpha = \theta(\beta, s_1)$. By Proposition 3.10, a function $\theta(\beta, s_1)$ for each fixed $\beta$ has exactly one minimum point $s_1^{\min}$ (see Fig. 10), and

$$\lim_{s_1 \to 1+0} \theta(\beta, s_1) = \lim_{s_1 \to +\infty} \theta(\beta, s_1) = +\infty.$$

Thus a function $s_1(\alpha)$ has two continuous branches $s_1 \in (1; s_1^{\min}]$ and $s_1 \in [s_1^{\min}; +\infty)$. The minimal root belongs to the left branch, therefore $\lim_{\alpha \to +\infty} s_1(\alpha) = 1 + 0$.

□

**Proposition 4.2.** *The right side of the equation (56) tends to $+\infty$ as $\alpha \to +\infty$.*

*Proof.* In the equation (56), substitute $\alpha = \dfrac{e^{\beta s_1} - 1}{e^{\beta}(s_1 - 1)}$:

$$\frac{\alpha^2}{2} e^{\beta} (s_1 - 1)^2 + \alpha s_1 = \frac{1}{2} e^{-\beta} \left( e^{\beta s_1} - 1 \right)^2 + \frac{s_1}{s_1 - 1} \cdot \frac{e^{\beta s_1} - 1}{e^{\beta}} \to +\infty \quad \text{as} \quad \alpha \to +\infty,$$

because $s_1 \to 1 + 0$ and, therefore, $\dfrac{s_1}{s_1 - 1} \to +\infty$, and other terms tend to nonzero constants

□

So, the left side of the equation (56) does not depend on $\alpha$, and the right side takes arbitrarily large values as $\alpha \to +\infty$.

On the other hand, the right side does not depend on $\Delta$, and the left side $\sim e^{\beta \Delta}$ as $\Delta \to +\infty$:

$$\frac{1}{A(\Delta)} \left( e^{\beta \Delta (m+1)} - 1 \right) = \frac{e^{\beta \Delta (m+1)} - 1}{e^{\beta} - 1 + \dfrac{e^{\beta \Delta (m+1)} - 1}{e^{\beta \Delta} - 1}} \sim e^{\beta \Delta}.$$

Assume that the left side of (56) is less than the right side for some $\Delta = \Delta_0$ (the parameters $\alpha, \beta$ are fixed). Then exists $\Delta > \Delta_0$ such that equality (56) holds.

Fix $\beta > 0$. Consider a condition (19). After a substitution $\alpha = \dfrac{e^{\beta s_1} - 1}{e^{\beta}(s_1 - 1)}$, we obtain:

$$e^{-\beta s_1} \left( e^{-\beta} + \frac{e^{\beta s_1} - 1}{e^{\beta}(s_1 - 1)} s_1 \right) > 1.$$

**Proposition 4.3.** *The function*

$$h(s_1) = e^{-\beta s_1} \left( e^{-\beta} + \frac{e^{\beta s_1} - 1}{e^{\beta}(s_1 - 1)} s_1 \right)$$

*is decreasing over $s_1 \geq 1$.*

*Proof.*

$$\frac{d}{ds_1} h(s_1) = -\frac{e^{-\beta(s_1+1)} \left( e^{\beta s_1} - \beta s_1 - 1 + \beta \right)}{(s_1 - 1)^2}.$$

Since $e^{\beta s_1} > 1 + \beta s_1$, we obtain $\frac{d}{ds_1} h(s_1) < 0$. Hence, $h$ is decreasing.

□

Therefore, if $s_1 = s_1^0 > 1$ satisfies (19), then all values $1 \leq s_1 \leq s_1^0$ also satisfy this condition.

Let $s_1 = 1 + \frac{1}{e^\beta}$, then

$$\alpha = \frac{e^{\beta s_1} - 1}{e^\beta (s_1 - 1)} = e^{\beta(1+e^{-\beta})} - 1.$$

Check that the condition (19) holds:

$$e^{-\beta s_1}(e^{-\beta} + \alpha s_1) = e^{-\beta(1+e^{-\beta})}\left(e^{-\beta} + (1 + e^{-\beta})\left(e^{\beta(1+e^{-\beta})} - 1\right)\right) =$$
$$= e^{-\beta(1+e^{-\beta})} e^{-\beta} + (1 + e^{-\beta}) - (1 + e^{-\beta}) e^{-\beta(1+e^{-\beta})} = 1 + e^{-\beta} - e^{-\beta(1+e^{-\beta})} > 1.$$

Because $s_1(\alpha)$ is decreasing, the condition (19) holds for each $\alpha \geq e^{\beta(1+e^{-\beta})} - 1$.

Now, we must check that the parameters $\alpha, \beta$ satisfy conditions (17) and (18) of Theorem 3.2.

**Proposition 4.4.** *If $\alpha > e^{\beta(1+e^{-\beta})} - 1$, then $\alpha > \varphi(\beta)$.*

*Proof.* By Proposition 3.10, $\theta(\beta, \tau_*) \geq \varphi(\beta)$ for each $\tau_*$. Let $\tau_* = 2$. Prove that $e^{\beta(1+e^{-\beta})} - 1 > \theta(\beta, 2)$.

$$e^{\beta(1+e^{-\beta})} - 1 > \frac{e^{2\beta} - 1}{e^\beta} \Leftrightarrow e^{\beta(2+e^{-\beta})} - e^\beta > e^{2\beta} - 1 \Leftrightarrow e^{\beta e^{-\beta}} > e^{-\beta} - e^{-2\beta} + 1.$$

After substitution $t = e^{-\beta}$, $t \in (0; 1)$, we obtain

$$e^{-t\ln t} > -t^2 + t + 1.$$

Expand the exponent into series. All terms of expansion for $t \in (0,1)$ are positive, so we can estimate it by first three terms (up to $t^2 \ln^2 t$):

$$e^{-t\ln t} = \sum_{n=0}^{\infty} (-1)^n \frac{t^n \ln^n t}{n!} > 1 - t\ln t + \frac{1}{2} t^2 \ln^2 t.$$

Now we can prove stronger inequality:

$$1 - t\ln t + \frac{1}{2} t^2 \ln^2 t > -t^2 + t + 1 \Leftrightarrow -\ln t + \frac{1}{2} \ln^2 t > -t + 1.$$

Substitute $t = e^{-\beta}$:

$$\beta + \frac{1}{2} e^{-\beta} \beta^2 > 1 - e^{-\beta} \Leftrightarrow e^{-\beta}\left(1 + \frac{\beta^2}{2}\right) > 1 - \beta.$$

For $\beta = 0$ there is an equality $1 = 1$. We will prove that the left side is increasing faster than the right side.

$$\frac{d}{d\beta}\beta + \frac{1}{2}e^{-\beta}\beta^2 > \frac{d}{d\beta}(1-\beta) \Leftrightarrow -\frac{1}{2}e^{-\beta}(\beta^2 - 2\beta + 2) > -1.$$

For $\beta = 0$ there is an equality $(-1) = (-1)$. The left side is increasing, then for $\beta > 0$ the inequality holds.

□

Fix $\alpha, \beta$ and let $\Delta = s_1(\alpha)$. Then the condition $\alpha \geq \theta(\beta, \Delta)$ holds (it turns into equality).

**Proposition 4.5.** For $s_1 = \Delta < 1 + \frac{1}{e^\beta}$ the equality holds: $\frac{1}{A(\Delta)}(e^{\beta\Delta(m+1)} - 1) \leq \frac{\alpha^2}{2}e^\beta(s_1 - 1)^2 + \alpha s_1$.

*Proof.* Estimate the left side:

$$\frac{1}{A(\Delta)}(e^{\beta\Delta(m+1)} - 1) = \frac{e^{\beta\Delta(m+1)} - 1}{e^\beta - 1 + \frac{e^{\beta\Delta(m+1)} - 1}{e^{\beta\Delta} - 1}} < e^{\beta\Delta} - 1.$$

We will prove that the right side is greater than $e^{\beta\Delta} - 1$:

$$\frac{\alpha^2}{2}e^\beta(\Delta - 1)^2 + \alpha\Delta = \frac{1}{2}e^{-\beta}(e^{\beta\Delta} - 1)^2 + \frac{\Delta}{\Delta - 1}\frac{e^{\beta\Delta} - 1}{e^\beta} > e^{\beta\Delta} - 1 \Leftrightarrow$$

$$\frac{1}{2}(e^{\beta\Delta} - 1) + \frac{\Delta}{\Delta - 1} > e^\beta \Leftrightarrow \frac{1}{2}(e^{\beta\Delta} - 1) + 1 + \frac{1}{\Delta - 1} > e^\beta.$$

Because $1 < \Delta < 1 + \frac{1}{e^\beta}$,

$$\frac{1}{2}(e^{\beta\Delta} - 1) + 1 + \frac{1}{\Delta - 1} \geq \frac{1}{2}(e^{\beta\Delta} - 1) + 1 + e^\beta > e^\beta.$$

□

**Lemma 4.6.** $\frac{W(-e^{-\beta-1}) + 1}{\beta} > \frac{1}{e^\beta}$.

*Proof.*

$$\frac{W(-e^{-\beta-1}) + 1}{\beta} > \frac{1}{e^\beta} \Leftrightarrow W(-e^{-\beta-1}) > \beta e^{-\beta} - 1 \Leftrightarrow -e^{-\beta-1} < (1 - \beta e^{-\beta})\exp(\beta e^{-\beta}) \Leftrightarrow$$

$$\exp(-\beta - \beta e^{-\beta}) < 1 - \beta e^{-\beta} \Leftrightarrow \exp(-\beta(1 + e^{-\beta})) < 1 - \beta e^{-\beta}.$$

Trivially, $\exp(-\beta(1 + e^{-\beta})) < e^{-\beta}$. Now we'll prove that $e^{-\beta} < 1 - \beta e^{-\beta}$:

$$e^{-\beta} < 1 - \beta e^{-\beta} \Leftrightarrow e^\beta > 1 + \beta,$$

which is true for each $\beta > 0$.

□

By Lemma 4.6, the function $\theta(\beta, \Delta)$ is decreasing over the interval $(1; \Delta_{min})$, where (see Proposition 3.10)

$$\Delta_{min} = 1 + \frac{W(-e^{-\beta-1}) + 1}{\beta} > 1 + \frac{1}{e^\beta}.$$

Hence, the condition $\alpha > \theta(\beta; \Delta)$ is also holds on that interval. If $\Delta \geq \Delta_{min}$, then we need to check the condition $\alpha \geq \varphi(\beta)$; it holds by Proposition 4.4.

Therefore, conditions of Lemma 3.7 remain true when $\Delta$ increasing. Then exists $\Delta$, satisfying equation (56).

Now we prove that the parameters $\alpha, \beta, \Delta$ determined above, satisfy the conditions of Lemma 3.13, or the conditions (20) – (22) of Theorem 3.2. Since $s_1 < \tau_k - \tau_{k-1} = \Delta$, we need to check the conditions (20) for $k \geq 2$:

$$\frac{\alpha^2}{2} e^\beta (s_1 - 1)^2 + \alpha s_1 > \frac{e^{\beta \tau_k} - 1}{\sum_{i=0}^{k-1} e^{\beta \tau_i}} = B_k, \quad k = 2, \ldots, m.$$

Since $\tau_k = \Delta k$ for $k = 2, \ldots, m$:

$$B_k = \frac{e^{\beta \tau_k} - 1}{\sum_{i=0}^{k-1} e^{\beta \tau_i}} = \frac{e^{\beta \Delta k} - 1}{e^\beta - 1 + \frac{e^{\beta \Delta k} - 1}{e^{\beta \Delta} - 1}} = \frac{1}{\frac{e^\beta - 1}{e^{\beta \Delta k} - 1} + \frac{1}{e^{\beta \Delta} - 1}}.$$

$B_k$ is increasing for $k \in \mathbb{N}$, so we must only check that the condition holds for $k = m$. For $k = m + 1$ the condition (20) turns into equality (56), then for $k \leq m$ inequality (20) holds.

Consider a case $k = 1$. If $s_1 > \tau_1 - \tau_0 = \Delta - 1$, then we must check the condition (21). We'll prove that it is weaker than the condition (19), that was checked above.

Write down the condition (21) for $k = 1$:

$$\frac{\alpha^2}{2} e^\beta (s_1 - 1)^2 + \alpha s_1 > \frac{e^{\beta(s_1+1)} - 1}{e^\beta} \Leftrightarrow \left(\frac{\alpha^2}{2} e^\beta (s_1 - 1)^2 + \alpha s_1 + e^{-\beta}\right) e^{-\beta s_1} > 1.$$

From (19),

$$(\alpha s_1 + e^{-\beta}) e^{-\beta s_1} > 1,$$

and a positive term $\frac{\alpha^2}{2} e^\beta (s_1 - 1)^2 e^{-\beta s_1}$ just make the inequality stronger.

If $s_1 \leq \Delta - 1$, then we can use for $k = 1$ the same arguments, that we used above for $k = 2, 3, \ldots, m$.

In the reason to verify (22):

$$\frac{\alpha^2}{2} e^\beta (s_1 - 1)^2 + \alpha s_1 > \frac{e^{\beta \tau_k}(e^{\beta s_1} - \alpha s_1) - 1}{\sum_{i=0}^{k-1} e^{\beta \tau_i}}, \quad k = 1, \ldots, m$$

we just check that

$$e^{\beta s_1} - \alpha s_1 \leq 1,$$

then these conditions are weaker than the conditions (20). Remark: we actually proved these conditions for each $1 \leq k \leq m$, but for $k = 1$, maybe, we are needed to check the condition (21) instead.

Make a substitution $e^{\beta s_1} - 1 = \alpha e^\beta(s_1 - 1)$.

$$\alpha e^\beta(s_1 - 1) - \alpha s_1 \leq 0 \Leftrightarrow e^\beta(s_1 - 1) - s_1 \leq 0 \Leftrightarrow s_1 < 1 + \frac{1}{e^\beta - 1}.$$

By construction, $s_1 < 1 + \frac{1}{e^\beta}$, therefore inequality holds.

As a result, we proved

**Theorem 4.7**. *For arbitrary $\beta > 0$, $\alpha \geq e^{\beta\left(1 + \frac{1}{e^\beta}\right)} - 1$ there exists $\Delta > 1$, such that the parameters $\alpha, \beta, \Delta$ satisfy the equation (55) and the conditions (17) – (22) of Theorem 3.2.*

## 5. CONCLUSION

We have introduced a fully coupled system of Mackey-Glass generators (3). After normalizations and replacements, system (3) was transformed into (4), for which we considered limit system (6). The existence of discrete traveling waves (7) was proved, and function $u(t)$ was found as a periodic solution of the equation (14). This solution contained the smallest possible number of switchings. Next, the solvability of the period equation (9) was proved, which implies the existence of a solution to system (6).

Figures 11, 12 and 13 show the results of a numerical experiment for systems (6) and (4). Note that, for $\gamma = 100$, the graphs of the solutions of the systems (4) and (6) are visually indistinguishable.

## REFERENCES

x

FUNDING

The work of V. V. Alekseev was supported by the Russian Science Foundation (project No. 21-71-30011).

The study of M. M. Preobrazhenskaia was funded by a grant Russian Science Foundation No. 22-11-00209, https://rscf.ru/project/22-11-00209/.

The work of V.K. Zelenova was carried out within the framework of a development programme for the Regional Scientific and Educational Mathematical Center of the Yaroslavl State University with financial support from the Ministry of Science and Higher Education of the Russian Federation (Agreement on provision of subsidy from the federal budget No. 075-02-2023-948).

CONFLICT OF INTEREST

The authors declare that they have no conflicts of interest.

FIGURE CAPTIONS

**Fig. 1.** A ring of $m$ generators.

**Fig. 2.** A fully coupled network of $m$ generators.

**Fig. 3.** The set $S$ of initial functions of the equation (14).

**Fig. 4.** The function $u_*(t)$.

**Fig. 5.** Relative positions of graphs of the functions $f(s)$, $g(s)$ on condition $s_1 \in [1, \tau_*]$. Left: $f(1) > g(1)$ and $f(\tau_*) < g(\tau_*)$. Right: $f(1) > g(1)$, $f(\tau_*) > g(\tau_*)$, but $f(s_0) < g(s_0)$ for $s_0 = argmin(f(s) - g(s))$, $s_0 \in [1; \tau_*]$.

**Fig. 6.** The function f(x) = xe$^x$ is decreasing over x < −1 and increasing over x > −1. The hatched area corresponds to eq. (40).

**Fig. 7.** The parameter domains, described in Lemma 3.7, for different $\tau_*$. There are functions θ(β, $\tau_*$), φ(β), δ(β, $\tau_*$), lines α = β and α = eβ represented on the graph.

**Fig. 8.** A schematic graph of the function $\tilde{u}(t)$.

**Fig. 9.** A schematic graph of functions $\tilde{w}(t)$ and $\tilde{v}(t)$. In this case $t_0 + \tau_1 \notin [t_1 + \tau_0, t_1 + \tau_1]$, then over $t \in (t_1 + \tau_0, t_1 + \tau_1]$ function $\tilde{v}(t)$ takes form (49) for $k = 1$. The sufficient condition that $\tilde{v}(t) > 1$ over this interval are (21) and (22) for $k = 1$. Over the next interval $[t_1 + \tau_1, t_1 + \tau_2]$, $t_0 + \tau_2 \in [t_1 + \tau_1, t_1 + \tau_2]$, so $\tilde{v}(t)$ defined by (50) for $k = 2$; the correspondent sufficient conditions that $\tilde{v}(t) > 1$ are (20) and (21) for $k = 2$.

**Fig. 10.** For each fixed $\beta$, the function $\theta(\beta, s_1)$ has exatly one minimum $s_1^{min}$, is decreasing over $(1; +s_1^{min}]$ and increasing over $[s_1^{min}; +\infty)$. $s_1(\alpha)$ is the minimal root of the equation $\alpha = \theta(\beta, s_1)$, therefore the left branch (thick line) defines $s_1(\alpha)$.

**Fig. 11.** A periodical solution of eq. (9), obtained numerically for the parameters $m = 5, \alpha = 10.0, \beta = 2.0, \Delta \approx 1.41$. The period is $(m + 1)\Delta \approx 8.45$.

**Fig. 12.** Dashed lines: a numerical solution of the limit system (6) with $m + 1 = 3$ generators for parameters $\alpha = 5.0, \beta = 1.2, \Delta \approx 1.801$. Solid lines: a numerical solution of system (4) for the same $m, \alpha, \beta$ and $\gamma = 10$.

**Fig. 13.** Pseudo-phase portraits of solutions of the limit system (6) and the correspondent system (4) with $\gamma = 10$ (left fig.) and $\gamma = 100$ (right fig.)

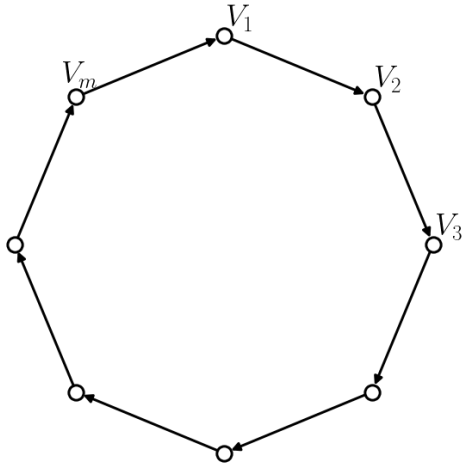

Fig. 1.

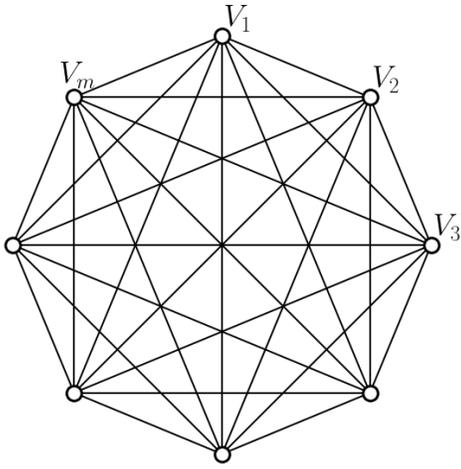

Fig. 2.

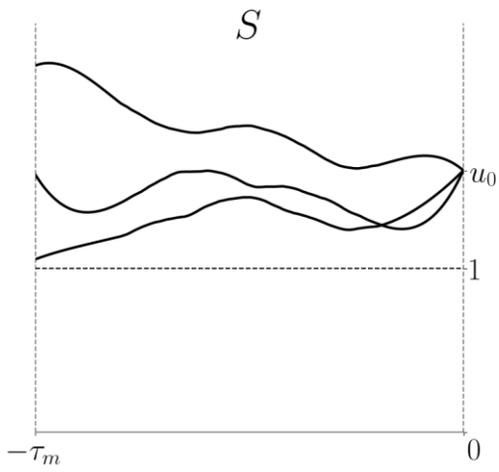

Fig. 3.

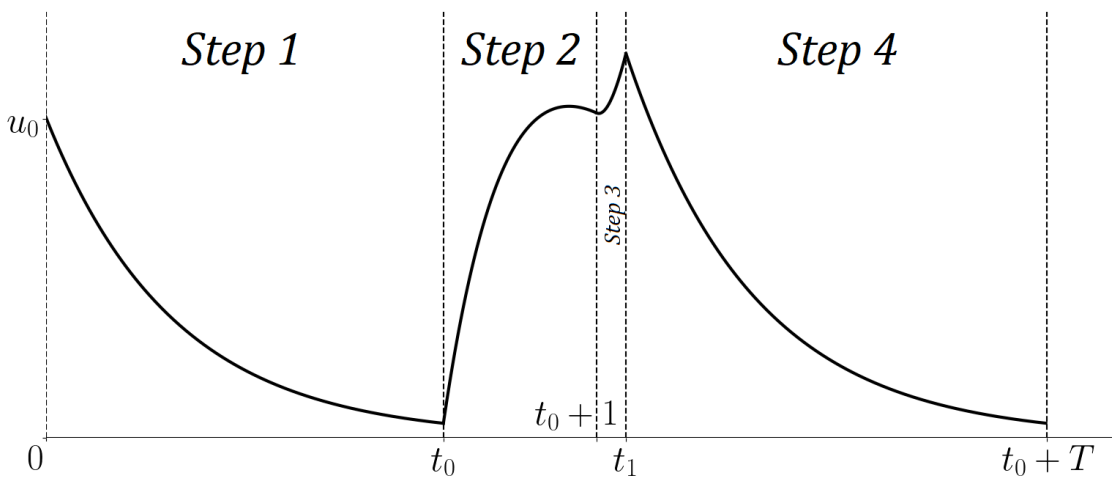

Fig. 4.

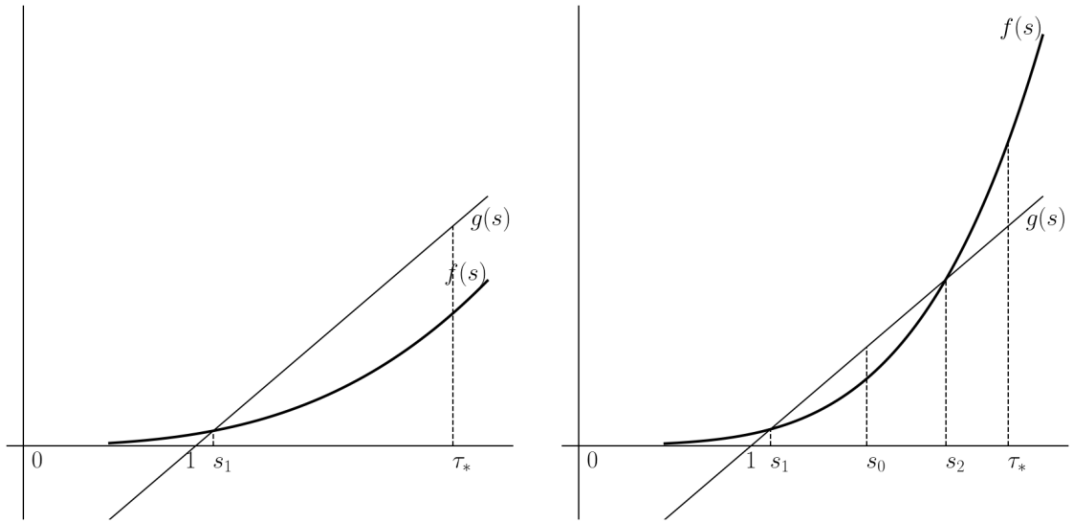

Fig. 5.

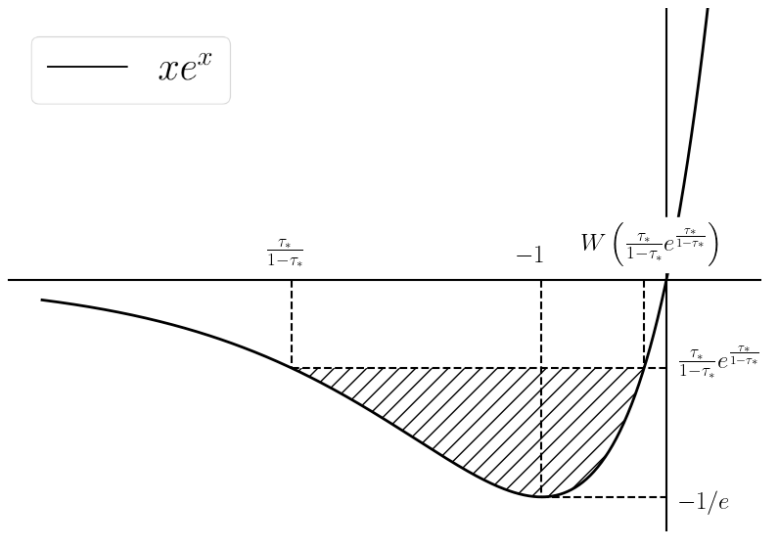

Fig. 6.

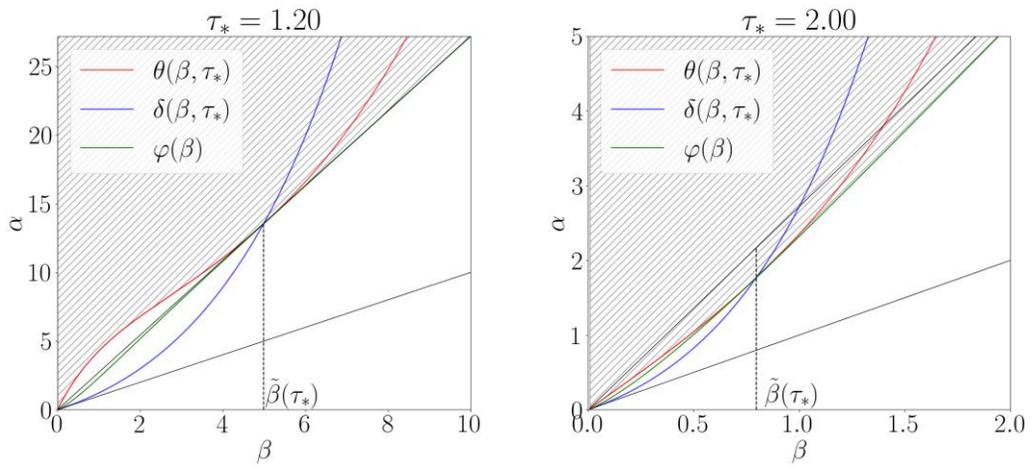

Fig. 7.

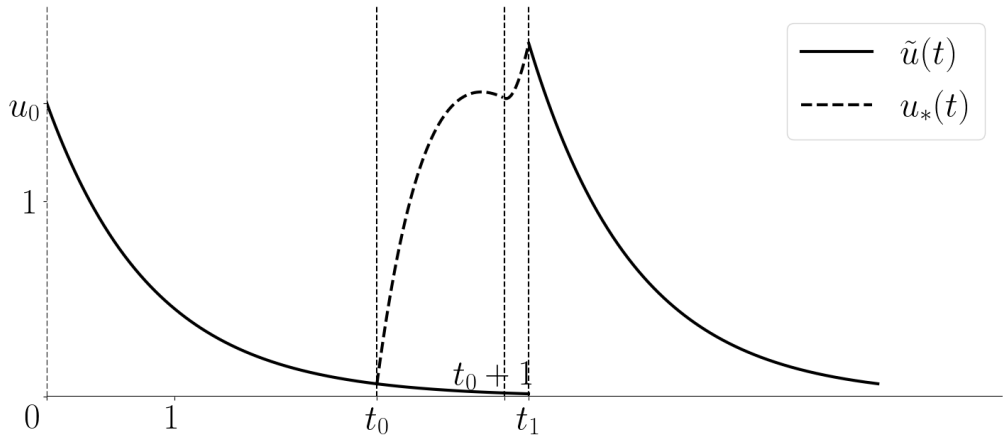

Fig. 8.

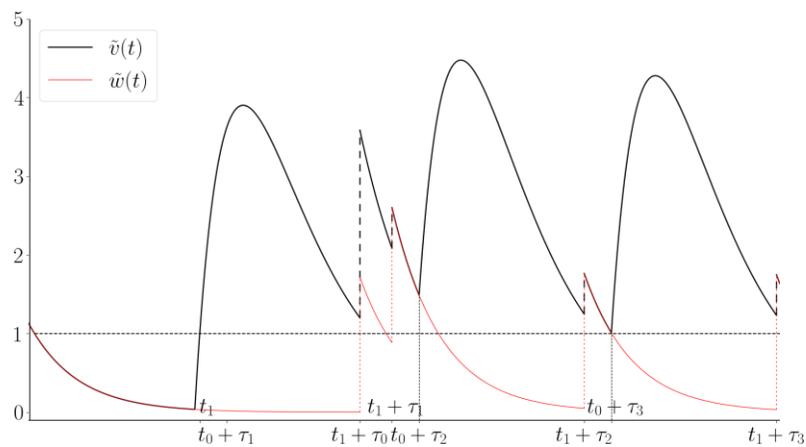

Fig. 9.

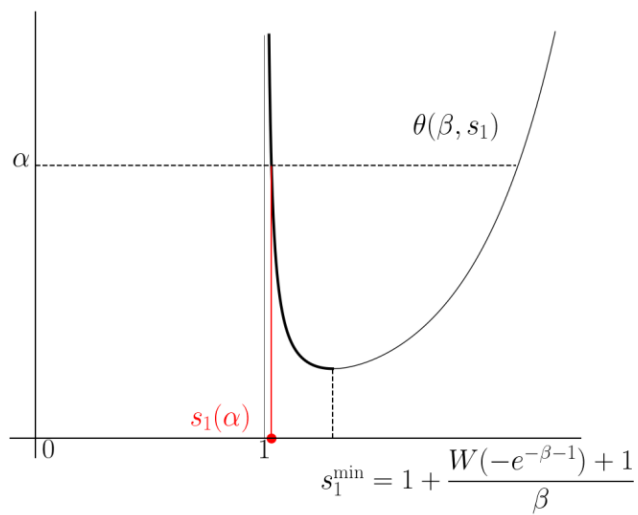

Fig. 10.

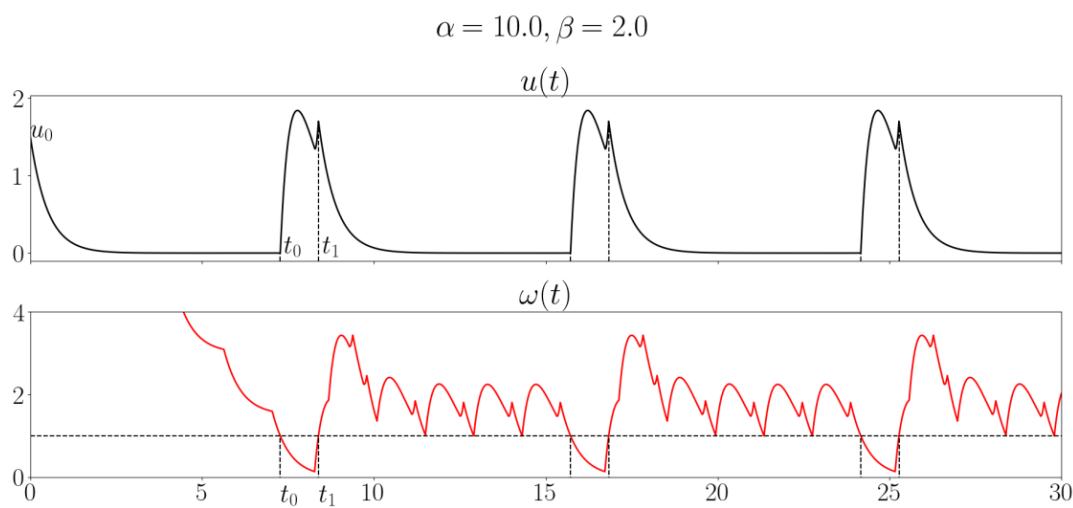

Fig. 11

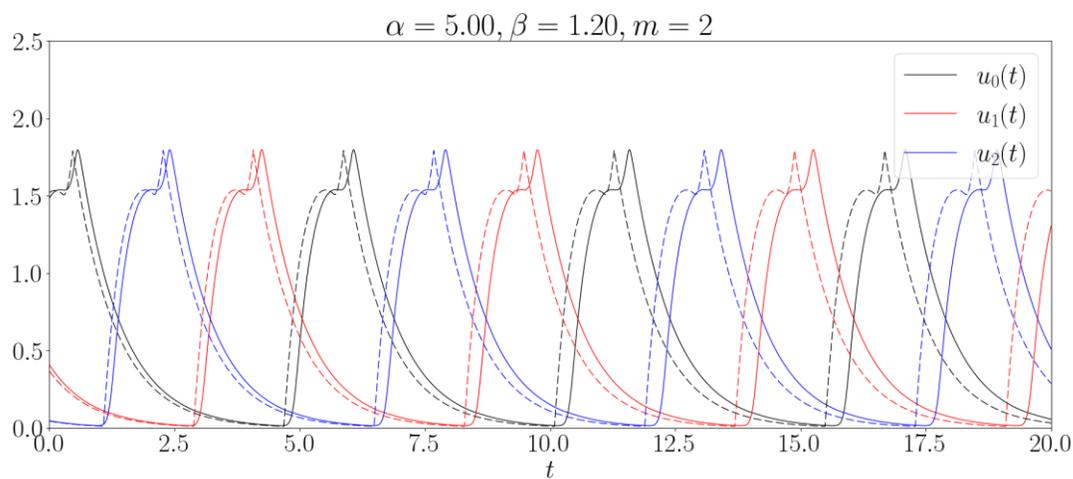

Fig. 12

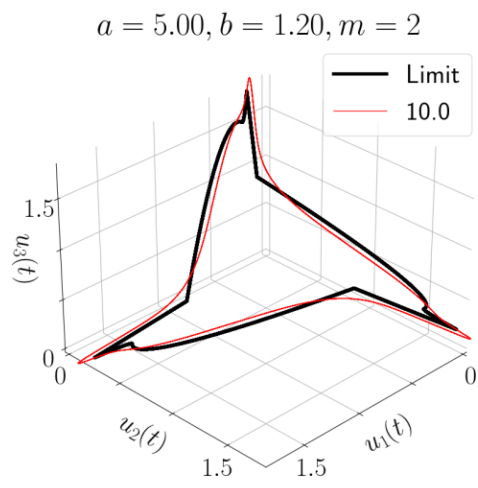 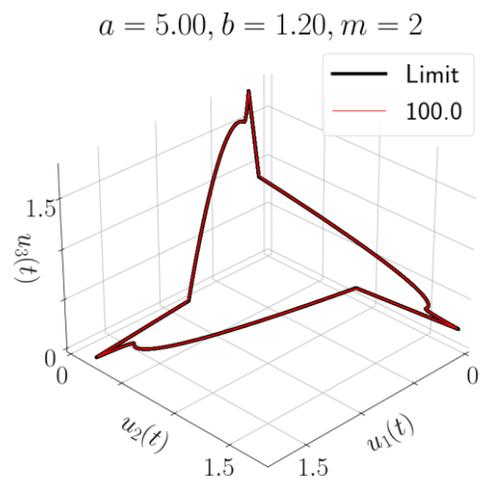

Fig. 13